\documentclass[]{aa}
\usepackage{psfig}
\begin{document}

\thesaurus{08(08.15.1)}

\title{Multiperiodicity in semiregular variables}
\subtitle{I. General properties}

\author{L.L. Kiss\inst{1} \and K. Szatm\'ary\inst{1} \and R.R. Cadmus, Jr.\inst{2}
\and J.A. Mattei\inst{3}}

\institute{Department of Experimental Physics and Astronomical Observatory,
JATE University,
Szeged, D\'om t\'er 9., H-6720 Hungary \and Department of Physics,
Grinnell College, Grinnell, IA 50112 USA \and
American Association of Variable Star Observers (AAVSO), 25 Birch
Street, Cambridge MA02138-1205 USA}

\titlerunning{Multiperiodicity in semiregulars}
\authorrunning{Kiss et al.}
\offprints{l.kiss@physx.u-szeged.hu}
\date{}

\maketitle
 
\begin{abstract}

We present a detailed period analysis for 93 red semiregular variables
by means of Fourier and wavelet analyses of long-term visual observations
carried out by amateur astronomers. The results of this analysis
yield insights into the mode structure of semiregular variables
and help to clarify the relationship between them and Mira variables.

After collecting all available data from various international databases
(AFOEV, VSOLJ, HAA/VSS and AAVSO) we test the accuracy and reliability of
data. We compare the averaged and noise-filtered visual light curves
with simultaneous photoelectric V-measurements, the effect
of the length versus the relatively low signal-to-noise ratio is
illustrated by period analysis of artificial data, while binning effects
are tested by comparing results of frequency analyses of the unbinned
and averaged light curves.

   The overwhelming majority of the stars studied show multiperiodic
behaviour. We found two significant periods in 44 variables, while
there are definite signs of three periods in 12 stars. 29 stars turned
out to be monoperiodic with small instabilities in the period.
Since this study deals with the general trends, we wanted to find
only the most dominant periods.

   The distribution of periods and period ratios is examined through the use of
the $({\rm log} P_0, {\rm log} P_1)$ and $({\rm log} P_1, {\rm log} P_0/P_1)$
plots. Three significant
and two less obvious sequences are present which could be explained
as the direct consequence of different pulsational modes. This
hypothesis is supported by the results for multiperiodic variables with
three periods. Finally, these space methods
are illustrated by several interesting case studies that show
the best examples of different special phenomena such as long-term
amplitude modulation, amplitude decrease and mode switching.

 
\end{abstract}

\section{Introduction}

Mira and semiregular variables (SRV's) are pulsating
low and intermediate mass red giants located on the asymptotic
giant branch (AGB). The importance of these variables is highlighted by
the fact that they are primary sources for the enrichment of
interstellar medium via mass loss.
The observed pulsational behaviour may lead to a better understanding
of inner physical processes having crucial effects on
stellar evolution.

The classification scheme according to the General Catalog of
Variable Stars (GCVS) is based only on the
amplitude and regularity of the visual variation. SRV's have amplitudes
smaller than
2.5 mag in V, while typical periods range from 25 to hundreds of days.
Their basic properties (classification, temperature, luminosity, space
distribution, important spectral features) were studied in general by
Kerschbaum \& Hron (1992), Jura \& Kleinmann (1992), Kerschbaum \&
Hron (1994), Lebzelter et al. (1995),
Kerschbaum \& Hron (1996), Kerschbaum et al. (1996), Hron et al. (1997).
Although they are usually treated separately from Mira-type variables,
there has been increasing evidence of a closer relationship between
the two types of variables. Kerschbaum \& Hron (1992, 1994) claimed
that some semiregulars are more closely related to Miras than the
pure classification suggests. Szatm\'ary et al. (1996) found V Boo to
have dramatically decreasing amplitude over decades of time mimicing
evolution from the Mira to the semiregular state. A similar phenomenon
was found by Bedding et al. (1998) for R Dor, which implies that
certain groups of semiregulars may belong to a subset of Mira variables.
Bedding \& Zijlstra (1998) reached similar conclusion based on HIPPARCOS
period-luminosity relations for Mira and semiregular variables.

The mode of pulsation in SRV's raised many questions during
the last decades. A detailed review is given by Percy \& Polano (1998),
who showed that the presence of higher overtone pulsation is suggested by
the
observations (up to the third and fourth overtone). Wood et al. (1998)
presented 5 different period-luminosity sequences for the LMC
red variables based on the MACHO photometric database,
concluding similarly to Percy \& Polano (1998) that even third and
fourth overtones could be the dominant excited modes.
Bedding et al. (1998) claim that the observed mode switching
in R Dor occurs between the first and the third overtone.
All of these studies support the idea that fundamental plus
first overtone pulsation in SRV's is an oversimplified
assumption and the complex light variations may be due to
many simultaneously excited modes.

As has been mentioned above, the typical time scale of SRV's can be
hundreds of days, and consequently there are very few high-quality
photometric observations (photographic or photoelectric) in the literature.
Although micro-lensing projects (MACHO, EROS, OGLE) yielded many
theoretical constraints on stellar pulsation interpretations
(see Welch 1998 for a review), the majority of SRV's need much
longer (a few decades, at least) continuous time-series of observations.
First results
concerning red variables in the LMC have already appeared
from the MACHO group (Cook et al. 1997, Minitti et al. 1998,
Alves et al. 1998, Wood et al. 1998), but periodicities in
SRV's in our own Galaxy deserve further study.
Fortunately, a large fraction of bright SRV's have been observed
visually by amateur astronomers all around the world. There exist
50--70 years long data series which are perfectly usable for
studying periodicities in the light curves (see e.g.
Percy et al. 1993, Mattei et al. 1998, Andronov 1998).

The main aims of this study are to present a detailed light curve
analysis for 93 SRV's based on long-term visual observations
and to demonstrate the
general trends and the most interesting phenomena we found
in the analysed sample.
The paper is organised as follows. Observations are discussed and
tested
in Sect.\ 2., while Sect.\ 3. deals with the results of
period analysis, especially with multiperiodicity as a
consequence of multimode pulsation.
Interesting special cases (triple periodicity, long-term amplitude
decrease and amplitude modulation) are briefly summarized in Sect.\ 4.

\begin{table*}
\small
\begin{center}
\caption{The list of programme stars. Variability types, periods and
spectral types are taken from the GCVS.}
\begin{tabular}{lllrllllrl}
\hline
GCVS       &  IRAS & Type  &  Period & Sp. type   &          GCVS  & IRAS     & Type  &  Period & Sp. type \\
\hline
{\bf O-rich} & & & & &                                                            Y UMa      & 12380+5607   &  SRb   &  168    &  M7II-III:\\
RU And     & --           &  SRa   &  238    &  M5e-M6e&                          Z UMa      & 11538+5808   &  SRb   &  196    &  M5IIIe\\
RV And     & 02078+4842   &  SRa   &  171    &  M4e&                              RY UMa     & 12180+6135   &  SRb   &  310    &  M2-M3IIIe\\
V Aqr      & 20443+0215   &  SRa   &  244    &  M6e&                              ST UMa     & 11251+4527   &  SRb   &  110    &  M4-M5III\\
S Aql      & 20093+1528   &  SRa   &  146    &  M3e-M5.5e&                        R UMi      & 16306+7223   &  SRb   &  326    &  M7IIIe\\
GY Aql     & 19474$-$0744 &  SR    &  204    &  M6III:e-M8&                       V UMi      & 13377+7433   &  SRb   &  72     &  M5IIIab\\
T Ari      & 02455+1718   &  SRa   &  317    &  M6e-M8e&                          SW Vir     & 13114$-$0232 &  SRb   &  150    &  M7III\\
RS Aur     & --           &  SRa   &  170    &  M4e-M6e&                          RU Vul     & --           &  SRa   &  174    &  M3e-M4e\\
U Boo      & 14520+1753   &  SRb   &  201    &  M4e&                                         &        &         &   \\
V Boo      & 14277+3904   &  SRa   &  258    &  M6e&                              {\bf C-rich} & & &\\
RV Boo     & 14371+3245   &  SRb   &  137    &  M5e-M7e&                          ST And     & 23362+3529   &  SRa   &  328    &  C4,3e-C6,4e\\
RS Cam     & 08439+7908   &  SRb   &  89     &  M4III&                            VX And     & 00172+4425   &  SRa   &  369    &  C4,5\\
RR Cam     & 05294+7225   &  SRa   &  124    &  M6 &                              AQ And     & 00248+3518   &  SR    &  346    &  C5,4\\
RY Cam     & 04261+6420   &  SRb   &  136    &  M3III&                            V Aql      & 19017$-$0545 &  SRb   &  353    &  C5,4-C6,4\\
RT Cnc     & 08555+1102   &  SRb   &  60     &  M5III&                            S Aur      & 05238+3406   &  SR    &  590    &  C4-5\\
V CVn      & 13172+4547   &  SRa   &  192    &  M4e-M6eIIIa:&                     UU Aur     & 06331+3829   &  SRb   &  234    &  C5,3-C7,4\\
SV Cas     & 23365+5159   &  SRa   &  265    &  M6,5&                             S Cam      & 05356+6846   &  SRa   &  327    &  C7,3e\\
AA Cas     & 01163+5604   &  Lb    &         &  M6III&                            U Cam      & 03374+6229   &  SRb   &         &  C3,9-C6,4e\\
SS Cep     & 03415+8010   &  SRb   &  90     &  M5III&                            ST Cam     & 04459+6804   &  SRb   &  300    &  C5,4\\
DM Cep     & 22073+7231   &  Lb    &         &  M4&                               T Cnc      & 08538+2002   &  SRb   &  482    &  C3,8-C5,5\\
RS CrB     & 15566+3609   &  SRa   &  322    &  M7  &                             X Cnc      & 08525+1725   &  SRb   &  195    &  C5,4\\
W Cyg      & 21341+4508   &  SRb   &  131    &  M4e-M6eIII&                       Y CVn      & 12427+4542   &  SRb   &  157    &  C5,4J\\
RU Cyg     & 21389+5405   &  SRa   &  233    &  M6e-M8e&                          RT Cap     & 20141$-$2128 &  SRb   &  393    &  C6,4\\
RZ Cyg     & 20502+4709   &  SRa   &  276    &  M7,0-M8,2ea&                      WZ Cas     & 23587+6004   &  SRb   &  186    &  C9,2\\
TZ Cyg     & 19147+5004   &  Lb    &         &  M6&                               RS Cyg     & 20115+3834   &  SRa   &  417    &  C8,2e\\
AB Cyg     & --           & SRb   &  520    &  M4IIIe&                            RV Cyg     & 21412+3747   &  SRb   &  263    &  C6,4e\\
AF Cyg     & 19287+4602   &  SRb   &  93     &  M5e-M7&                           TT Cyg     & 19390+3229   &  SRb   &  118    &  C5,4e\\
U Del      & 20431+1754   &  SRb   &  110    &  M5II-III&                         AW Cyg     & 19272+4556   &  SRb   &  340    &  C4,5\\
CT Del     & 20270+0943   &  Lb    &         &  M7&                               V460 Cyg   & 21399+3516   &  SRb   &  180    &  C6,4\\
CZ Del     & 20312+0920   &  SRb   &  123    &  M5&                               RY Dra     & 12544+6615   &  SRb:  &  200    &  C4,5\\
EU Del     & 20356+1805   &  SRb   &  60     &  M6,4III&                          UX Dra     & 19233+7627   &  SRa:  &  168    &  C7,3\\
S Dra      & 16418+5459   &  SRb   &  136    &  M7&                               RR Her     & 16028+5038   &  SRb   &  240    &  C5,7e-C8,1e\\
TX Dra     & 16342+6034   &  SRb   &  78     &  M4e-M5&                           U Hya      & 10350$-$1307 &  SRb   &  450    &  C6,5\\
AH Dra     & 16473+5753   &  SRb   &  158    &  M7&                               V Hya      & 10491$-$2059 &  SRa   &  531    &  C6,3e-C7,5e\\
SW Gem     & 06564+2606   &  SRa   &  680    &  M5III&                            W Ori      & 05028+0106   &  SRb   &  212    &  C5,4\\
X Her      & 16011+4722   &  SRb   &  95     &  M6e&                              Y Per      & 03242+4400   &  M     &  249    &  C4,3e\\
ST Her     & 15492+4837   &  SRb   &  148    &  M6-7IIIas&                        SY Per     & 04127+5030   &  SRa   &  474    &  C6,4e\\
UW Her     & 17126+3625   &  SRb   &  104    &  M5e&                              S Sct      & 18476$-$0758 &  SRb   &  148    &  C6,4\\
g Her      & 16269+4159   &  SRb   &  89     &  M6III&                            Y Tau      & 05426+2046   &  SRb   &  242    &  C6.5,4e\\
RT Hya     & 08272$-$0609 &  SRb   &  290    &  M6e-M8e&                          SS Vir     & 12226+0102   &  SRa   &  364    &  C6,3e\\
RY Leo     & 10015+1413   &  SRb   &  155    &  M2e&                                         &        &         &       \\
U LMi      & 09516+3619   &  SRa   &  272    &  M6e&                              {\bf Uncertain} & & & & \\
RX Lep     & 05090$-$1155 &  SRb   &  60     &  M6,2III&                          T Cen      & 13388$-$3320 &  SRa   &  90     &  K0:e-M4II:e\\
SV Lyn     & 08003+3629   &  SRb   &  70     &  M5III&                            AI Cyg     & 20297+3221   &  SRb   &  197    &  M6-M7\\
X Mon      & 06548$-$0859 &  SRa   &  156    &  M1eIII-M6ep&                      GY Cyg     & --           &  SRb   &  300    &  M7p\\
BQ Ori     & 05540+2250   &  SR    &  110    &  M5IIIe-M8III&                     V930 Cyg   & 19371+3021   &  Lb    &         &      \\
UZ Per     & 03170+3150   &  SRb   &  927    &  M5II-III&                         RS Gem     & 06584+3035   &  SRb   &  140    &  M3-M8\\
$\tau^4$ Ser & 15341+1515 & SRb  &  100    &  M6IIb-IIIa&                         RX UMa     & 09100+6728   &  SRb   &  195    &  M5\\
W Tau      & 04250+1555   &  SRb   &  265    &  M4-M6.5& & & & &\\
V UMa      & 09047+5118   &  SRb   &  208    &  M5-M6& & & & &\\
\hline
\end{tabular}
\end{center}
\end{table*}

\section{Observations}

The bulk of the analysed data were taken from three
international databases of visual observations.
These belong to the Association Francaise des Observateurs d'Etoiles
Variables (AFOEV\footnote{\tt ftp://cdsarc.u-strasbg.fr/pub/afoev}),
the Variable Star Observers' League in Japan
(VSOLJ\footnote{\tt http://www.kusastro.kyoto-u.ac.jp/vsnet/gcvs})
and the Hungarian Astronomical Association -- Variable Star Section
(HAA/VSS\footnote{\tt http://www.mcse.hu/vcssz/data}).
The compiled visual estimates are stored at publicly
available web-sites as Julian Date + magnitude files.
A smaller fraction of data originated from the American
Association of Variable Star Observers (AAVSO International
Database which includes HAA/VSS, and part of AFOEV and
VSOLJ observations).

\begin{figure}
\begin{center}
\leavevmode
\psfig{figure=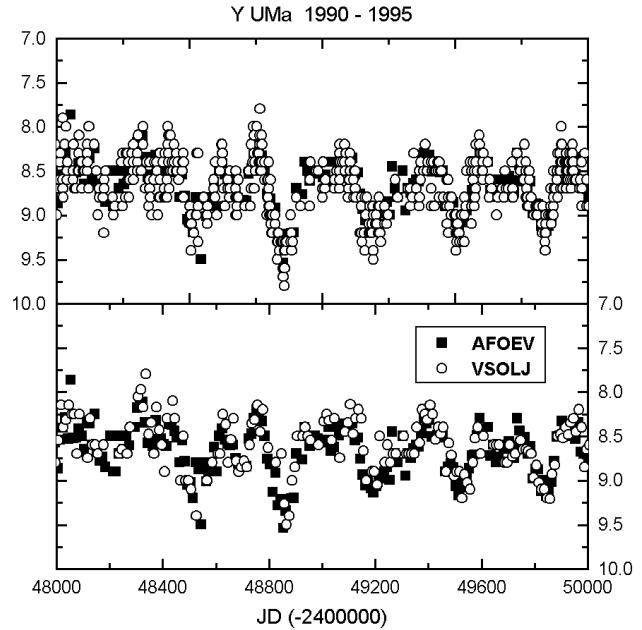,width=\linewidth}
\caption{A comparison of the visual light curves for Y UMa from
AFOEV and VSOLJ.
The differences do not exceed the uncertainty of the individual points
which is about $\pm0.3$ mag. The top panel shows the original data, while
the averaged curves are plotted in the bottom panel.}
\end{center}
\label{f1}
\end{figure}

The main selection criterion in choosing the sample was the length
and the continuity of the light curves. In order to reach high resolution
in the frequency domain, we usually kept only those stars with at least 10 years
of continuous data. This is equivalent to a frequency resolution
($\sim$time$^{-1}$) of {$2.7\cdot10^{-4}$} cycles/day. In most cases
the length of the analysed data is about 50 years, and occasionally it is
70--80 years. The final sample containing 93 semiregular variables
is summarized in Table 1, with the main information taken from the GCVS.
Y~Per (classified as a Mira star in the GCVS) was included because of its
recently observed semiregular nature (see Sect.\ 4.2.). A few
Lb-type variables are also included, as their classification is a
quite uncertain issue; recent studies of Kerschbaum et al. (1996)
and Kerschbaum \& Olofsson (1998) pointed out the close similarity
of selected Lb's and SRV's based on the infrared and mass-loss properties.

There are two steps in the data handling that precede
before the period analysis: (1) data averaging using 10-day bins, and
(2) merging of observations of different origin. We performed a few
simple tests
to decide whether merging should precede or follow the averaging. We plotted
the different original light curves together for the best observed
stars and found that the systematic differences did not exceed the level
of the scatter in the data. One example can be seen in Fig.\ 1, where we
have plotted
the French and Japanese data for Y~UMa (type SRb). The error of
an individual point is estimated to be about $\pm0.3$ mag.
The two curves are very similar,
which suggests that the comparison sequences define
a well determined system of visual magnitudes. A similar conclusion was drawn
for the majority of the stars in our sample, so we simply merged the
available data before calculating the averaged curves.
We could possibly reach somewhat better precision by introducing personal
corrections for the most active observers, but as other tests have
shown, the length of data is much more important in determining
periodicities -- which is our main goal -- than is the accuracy of
the individual measurements (see below). A thorough review of the
homogeneity of visual photometry is given by Sterken \& Manfroid (1992).

The averaging procedure consisted of taking 10-day bins and calculating
the mean value from the individual points. Since the typical time scale
of the period
in our sample of semiregular variables is about one hundred days,
this binning
procedure does not smooth out significant detail in the light variations.
A rough estimate
of the resulting improvement in
precision is as follows. As we mentioned earlier, the error of an
individual observation is about $\pm0.3$ mag. For a given 10-day bin with
10 points within it, the standard error of the mean value will
be $0.3/\sqrt{10}\approx0.1$ mag. The amount of data and their
distribution in our sample in most cases permit such precision
to be realized. Extremely deviant points differing from the mean value by
more than $\sim 3 \sigma$ were rejected in the original data
during a close visual inspection of all light curves.

\subsection{Tests of the quality and usability}

\begin{figure*}
\begin{center}
\leavevmode
\psfig{figure=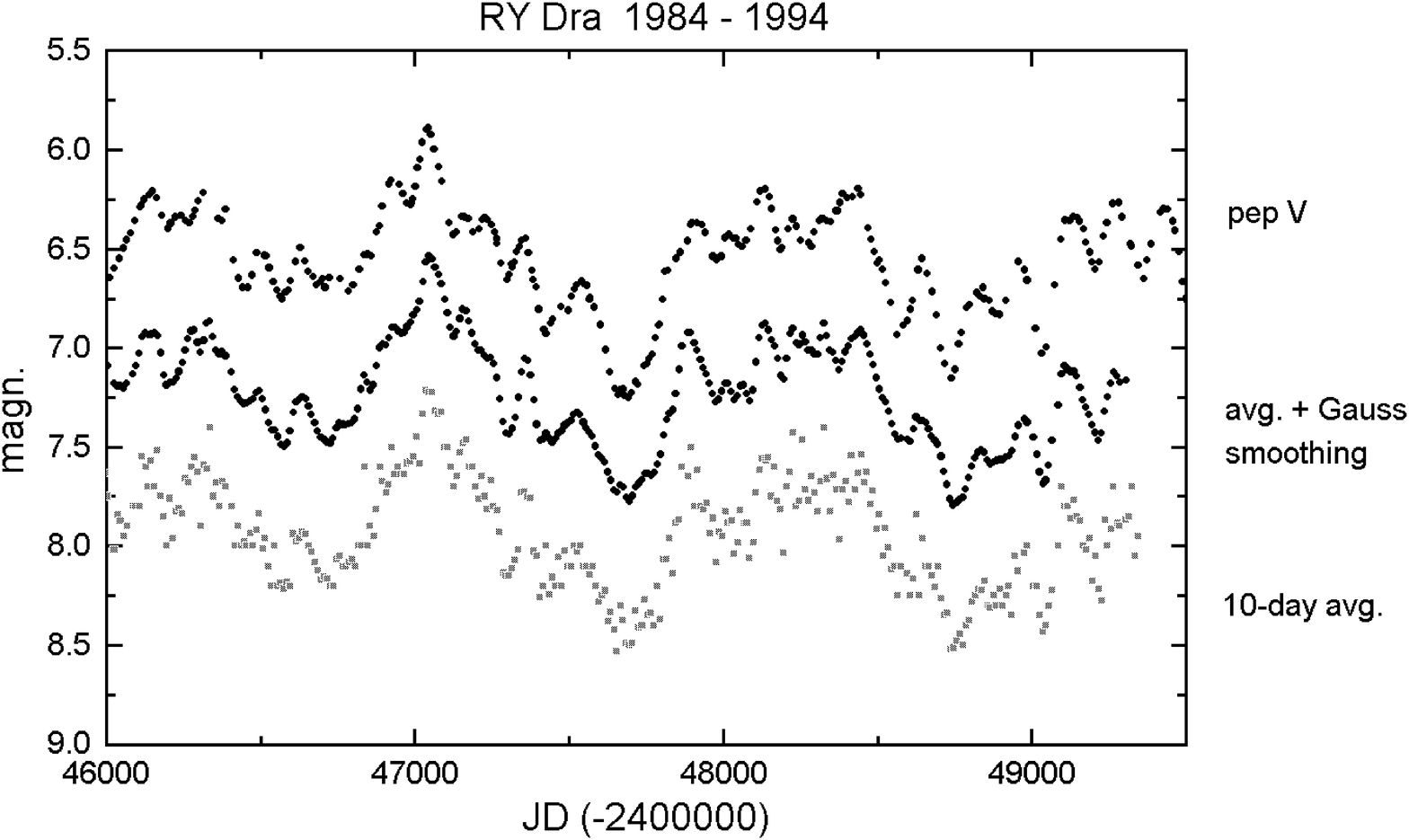,width=14cm}
\caption{Comparison between the photoelectric and visual observations
for RY~Dra.}
\end{center}
\label{f2}
\end{figure*}

\begin{figure*}
\begin{center}
\leavevmode
\psfig{figure=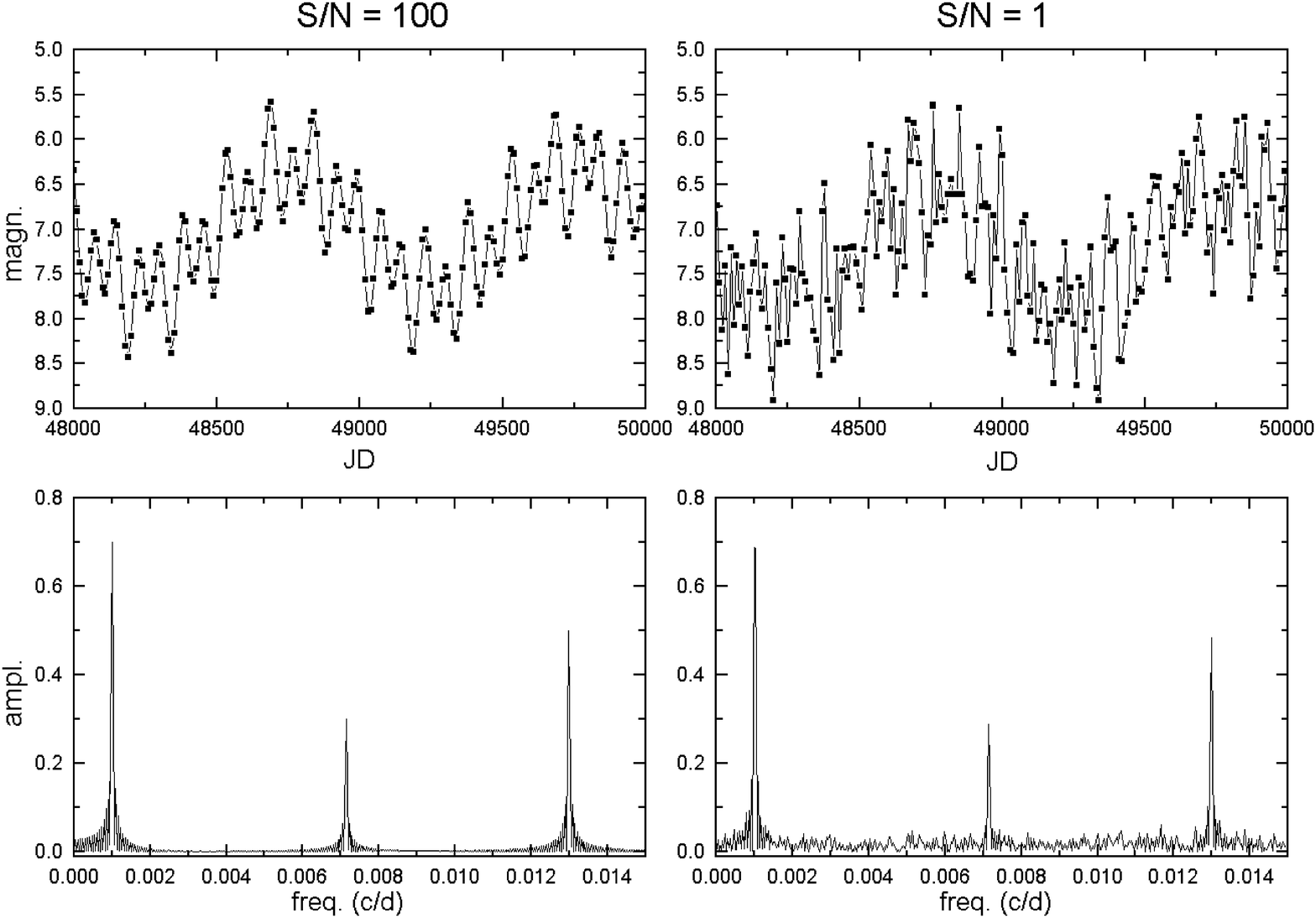,width=14cm}
\caption{Artifical data with different S/N ratios and their corresponding
DFT spectra. It can be seen that length is much more important that
the quality of data. Note that top panels show only subsets.}
\end{center}
\label{f3}
\end{figure*}

Three tests were made to check the reliability and usability
of the resulting
mean light curves. The first was comparison with available
simultaneous photoelectric V-measurements. Although the spectral
response function of the human eye differs from the Johnson
V filter, there were and continiue to be several attempts to
find calibrations of transformations.
Zissell (1998) modified the zeropoint of the conversion formula proposed
by Stanton (1981) and gave the following relation:
\begin{center}
$$m_{vis.}~=~ V~+~0.182~(B-V)~-~0.032. \eqno (1)$$
\end{center}
\noindent According to the available
photoelectric observations of semiregular variables, in most cases their
$B-V$ colour changes with much smaller amplitude than V brightness does,
thus it is a straightforward simplifying assumption that there is
a constant shift between the photoelectric V and visual light curves.
This is, of course, true only at a level of about 0.1 mag, which is
in the range of the scatter of the visual data.

A direct comparison is shown in Fig.\ 2, where we compare visual data
for RY~Dra with simultaneous photoelectric V measurements carried
out at Grinnel College. The top curve is the
photoelectric one, while the bottom curve is the corresponding
10-day mean of visual data. The middle curve is a noise filtered
version of the lower curve, where noise filtering was done by
a simple Gaussian smoothing with 8 days FWHM. Note that while
the visual curves were shifted in a vertical direction for clarity,
the distance between the smoothed visual and the photoelectric curve
is the real difference caused by the colour effects.
The observed average shift of 0.60 mag is in good agreement
with the predicted 0.57 mag by Eq.\ 1 ($\langle B-V \rangle\approx~3.3$
for RY~Dra).
The agreement between
the visual and photoelectric curves are very good, even the smallest
humps and bumps, of 0.1 mag are clearly visible in the
visual data. A similar conclusion can be drawn using Hipparcos Tycho V
data (ESA 1997\footnote{\tt http://astro.estec.esa.nl/Hipparcos}):
visual observations define light curves that are very similar
to the photoelectric ones. We have to note, that the Gaussian data
smoothing was applied only here because of its illustrative power,
our main analyses were based on the 10-day binned light curves only.

Another test was performed as a numerical simulation in order
to study the effect of the length of the data set versus the
signal-to-noise ratio (S/N).
We generated artificial time-series by adding three monoperiodic
signals with very similar periods and amplitudes to those
observed in real variables (e.g. A$_0=0.7$ mag, P$_0=1000$ days,
A$_1=0.3$ mag, P$_1=140$ days, A$_2=0.5$ mag, P$_2=77$ days).
Additional white noise was added to get artificial S/N values of 100
and 1, respectively. The ``observed'' time ranged from JD 2435000 to 2451000.
We calculated the Discrete Fourier Transforms (DFT) for both datasets
and the results are shown in Fig.\ 3.

It is obvious that the period (and amplitude) determination is
almost completely independent of the S/N ratio if the time-series
is long enough.
This can be explained by the fact that the applied noise
is independent of the current brightness and consequently
these two quantities are also independent in the frequency domain.
Real observations
come from many different observers who made their estimates
independently, therefore the observational noise is uncorrelated.
We have extensively
explored this question and our conclusion is that the analysed time-series
fulfill all requirements for accurate period analysis. This result is similar
to that of Szatm\'ary \& Vink\'o (1992). We have to
note that the independence of noise and observations can be
assumed only for bright semiregulars with amplitudes
that are not too large.
For Mira variables, which can become quite faint at minimum light,
the data obtained by observers using small telescopes will
have more scatter near the minima in the light curves.

Following the referee's note on using the averaged data, we have
performed a third test addressed to the effects of the binning. The most
important effect is the decrease of amplitude due to the binning, while
the resulting frequencies may differ a bit, too.
We explored this question by analysing the unbinned, 5-day and 10-day mean
light curves.

\begin{figure}
\begin{center}
\leavevmode
\psfig{figure=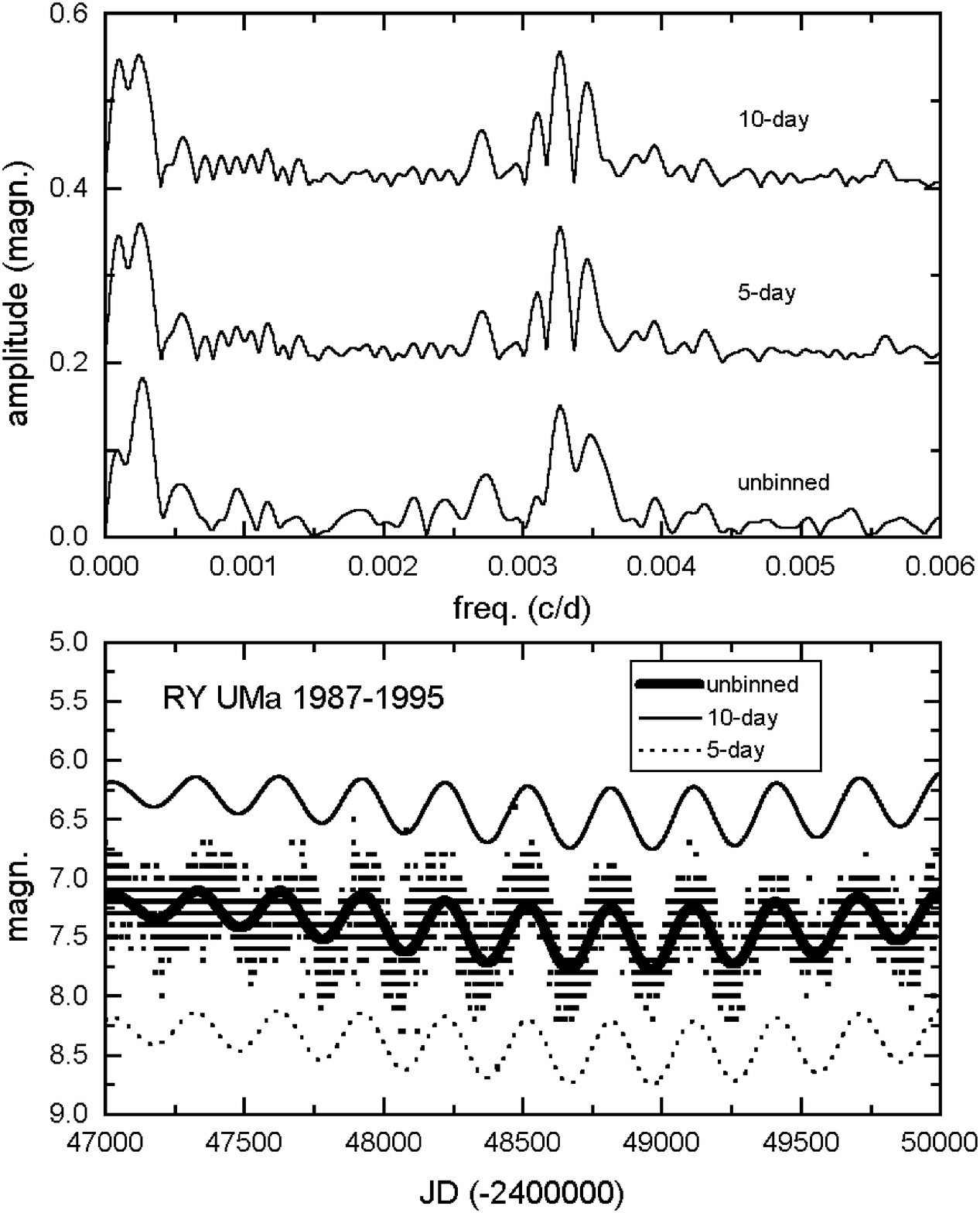,width=\linewidth}
\caption{The effects of binning for RY~UMa.
{\it Top panel}: comparison of the different Fourier-spectra
calculated from the original data and 5-day, 10-day bins.
{\it Bottom panel}: comparison of the four-component fitted curves
with the observations (small dots). A vertical shift
of $\pm$1 mag was applied for clarity.}
\end{center}
\label{f4}
\end{figure}

\begin{table}
\begin{center}
\caption{The frequencies and amplitudes of four principal peaks in the
Fourier-spectra plotted in Fig.\ 4.}
\begin{tabular} {llll}
\hline
      & unbinned    & 5-day      & 10-day\\
\hline
f$_0$ & 0.000104    & 0.000100 & 0.000100\\
A$_0$ & 0.119       & 0.146    & 0.152\\
f$_1$ & 0.000272    & 0.000252 & 0.000248\\
A$_1$ & 0.178       & 0.170    & 0.165\\
f$_2$ & 0.003276    & 0.003276 & 0.003271\\
A$_2$ & 0.145       & 0.153    & 0.156\\
f$_3$ & 0.003500    & 0.003480 & 0.003480\\
A$_3$ & 0.119       & 0.108    & 0.113\\
\hline
\end{tabular}
\end{center}
\end{table}

The results of this test are briefly summarized in Fig.\ 4 with those
of obtained for RY~UMa (type SRb). We plotted three different Fourier-spectra
calculated from the binned and original, unbinned data.
The principal peaks have slightly different frequencies and amplitudes
(Table 2),
but the four-component fits (bottom panel in Fig.\ 4) do not differ
significantly. This suggests that the differences are mainly due to the
uncertainty of the whole analysis caused by the noisy data and not
particularly due to the
averaging. We conclude that the averaging procedure does not
introduce significant alias structures, if the light curves are
densely covered by many independent observations. We obtained similar
results even for stars with periods of about 100 days (e.g. TX~Dra) suggesting
that data binning does not affect too seriously the calculated periods.
The amplitudes have, of course, larger uncertainties, but as they
may have cycle-to-cycle changes, this aspect is beyond our present scope.

\section{Period analysis}

We calculated Discrete Fourier Transforms (DFT) of the merged and
averaged time-series. The frequency ranged from 0 to 0.025 cycles/day,
while the frequency step was chosen as $4\cdot10^{-6}$ c/d.
The code used was Period98 (Sperl 1998\footnote{\tt
http://dsn.astro.univie.ac.at/period98}).
A few sample power spectra are presented in Figs.\ 5-6 and
Sect.\ 4.

\begin{figure*}
\begin{center}
\leavevmode
\psfig{figure=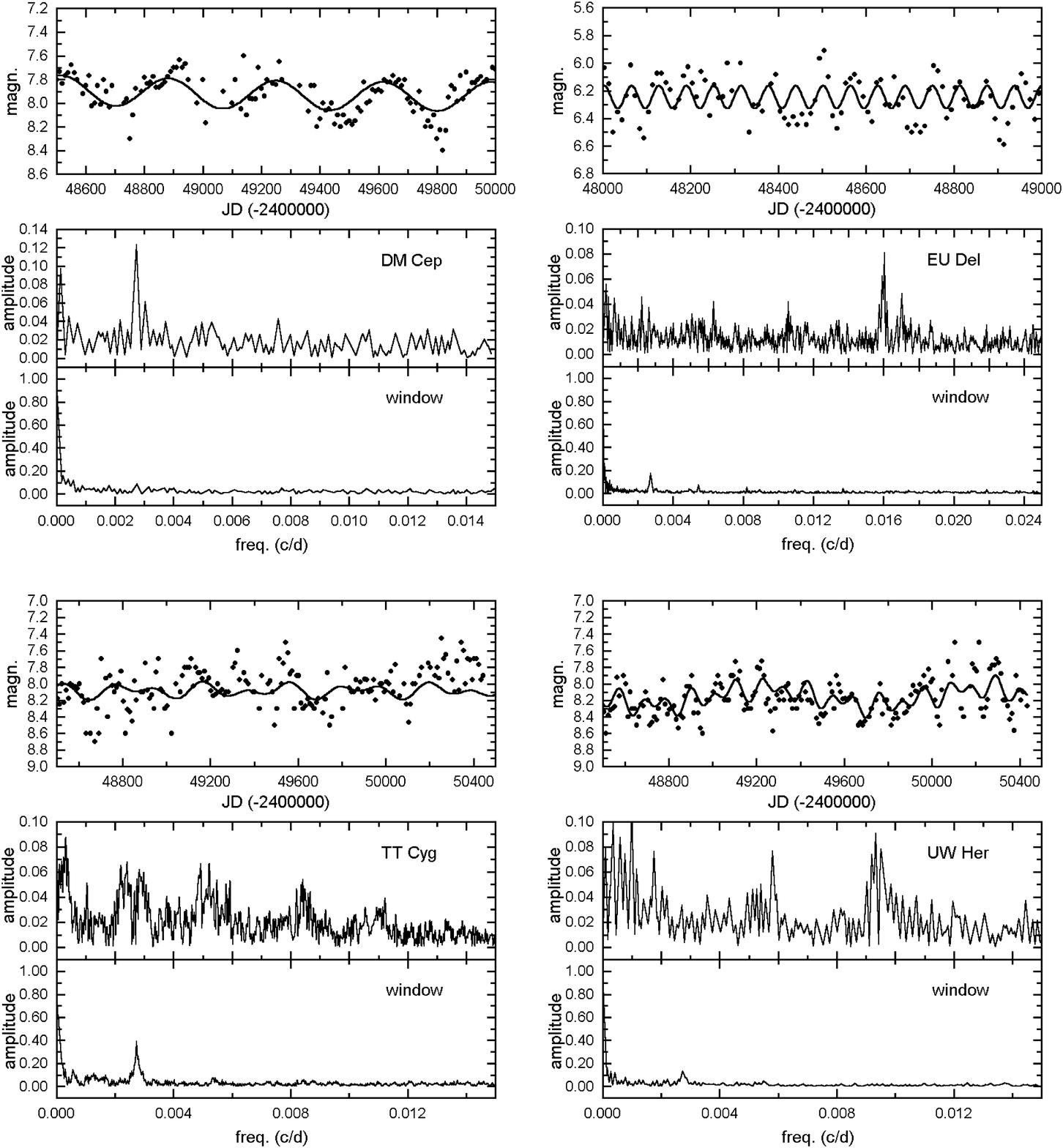,width=\textwidth}
\caption{Sample spectra of low-amplitude variables
with one- (DM~Cep and EU~Del), two- (TT~Cyg) and three-component
(UW~Her) fits. The averaged light curves were calculated with 10-day bins.}
\end{center}
\label{f5}
\end{figure*}

\begin{figure*}
\begin{center}
\leavevmode
\psfig{figure=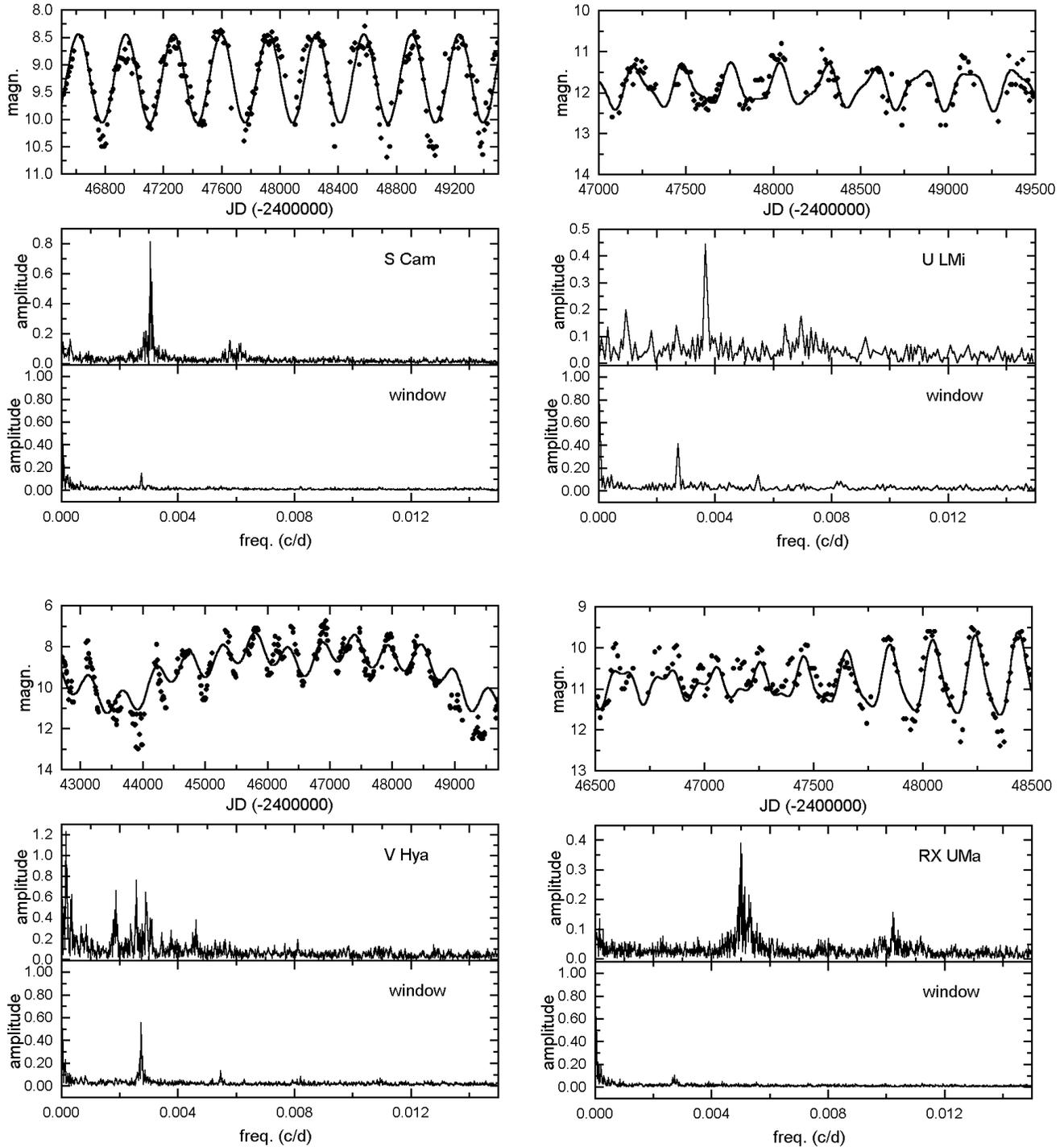,width=\textwidth}
\caption{Sample spectra of medium amplitude variables
with one- (S~Cam), two- (U~LMi and V~Hya) and three-component
(RX~UMa) fits. The averaged light curves were calculated with
10-day bins.}
\end{center}
\label{f6}
\end{figure*}

\begin{table}
\begin{center}
\caption{Monoperiodic variables. $\langle m \rangle$ is the
mean visual brightness, $\Delta T$ is the length of the time-series.
The numbers in brackets denote
the estimated uncertainty based on the width of the peaks in the
Fourier spectra. $A$ means the semi-amplitude.}
\begin{tabular} {lrrlll}
\hline
Star    & $\langle m \rangle$ &  $\Delta T$   & P (GCVS) & P  & $A$\\
\hline
RV And  & 10.0   &  22300     & 171   & 165(5)  & 0.30\\
S Aql   & 10.5   &  25800     & 146   & 143(3)  & 0.98\\
GY Aql  & 12.3   &   3000     & 204   & 464(4)  & 2.35\\
T Ari   & 9.5    &  33000     & 317   & 320(3)  & 0.91\\
S Aur   & 11.2   &  18800     & 590   & 596(6)  & 0.61\\
U Boo   & 11.2   &  23300     & 201   & 204(3)  & 0.62\\
RV Boo  & 8.5    &   7000     & 137   & 144(2)  & 0.09\\
S Cam   & 9.3    &  28000     & 327   & 327(1)  & 0.81\\
RY Cam  & 8.4    &   8500     & 136   & 134(1)  & 0.16\\
T Cnc   & 9.0    &  12400     & 482   & 488(4)  & 0.34\\
RT Cap  & 7.5    &  24000     & 393   & 400(4)  & 0.31\\
T Cen   & 7.0    &  12400     & 90   & 91(1)   & 0.62\\
DM Cep  & 7.9    &   6500     & --   & 367(3)  & 0.12\\
RS CrB  & 7.5    &  12400     & 322   & 331(1) & 0.19\\
AI Cyg  & 9.0    &   3000     & 197   & 146(2) & 0.18\\
GY Cyg  & 10.6   &   8000     & 300   & 143(1) & 0.13\\
V460 Cyg & 6.5   &   6400     & 180   & 160(10) & 0.08\\
V930 Cyg & 12.5  &   2000     & --   & 247(3) & 0.72\\
EU Del  &  6.2   &  11000     & 60   & 62(1)  & 0.08\\
SW Gem  &  8.8   &  11800     & 680   & 700(10) & 0.10\\
RR Her  &  9.0   &  27800     & 240   & 250(10) & 0.54\\
RT Hya  &  8.1   &  12400     & 290   & 255(3) & 0.20\\
U Hya   &  5.3   &  27800     & 450   & 791(5) & 0.06\\
X Mon   &  8.4   &  26600     & 156   & 148(7) & 0.59\\
SY Per  &  10.7  &  5000      & 474   & 477(9) & 0.89\\
UZ Per  &  8.7   &  4000      & 927   & 850(10) & 0.25\\
W Tau   &  10.4  &  24800     & 265   & 243(3) & 0.27\\
V UMa   &  10.4  &  12400     & 208   & 198(2) & 0.19\\
SS Vir  &  8.3   &  25500     & 364   & 361(1) & 0.81\\
\hline
\end{tabular}
\end{center}
There are some episodes in the light variation of 
RV~And, S~Aql and U~Boo caused by possible mode
switching (Cadmus et al. 1991).
\end{table}

We did not try to extract as many periods as possible from the
power spectra because the excited frequencies in semiregular variables
are not stable over time (see, e.g., Mattei et al. 1998).
The DFT may contain many misleading peaks
because of the cycle-to-cycle variations. These changes make
impossible to fit simple sums of sines. Our approach was to accept only
the most dominant periods which were
tested by whitening, cleaning and alias filtering. One-year
alias peaks occur for many stars, while in some cases cross production
terms are present as well (e.g. $f_0, f_1, f_0\pm f_1$). We did
an iterative period determination allowed by Period98 (Sperl 1998), in which
we checked the consistency of the fitted harmonics and the light
curve itself after every step. The frequency identification was a
quite difficult task in certain stars (e.g. TT~Cyg, TZ~Cyg and other
low-amplitude variables), mainly because of the long-term
changes in the mean brightness. They may cause fairly high false
peaks in the low-frequency region which have to be subtracted and
neglected in searching for pulsational periods. This involves
some additional uncertainty which can be hardly avoided.
The instabilities of the excited frequencies may cause multiple
peaks scattering around local average values in the Fourier spectra,
therefore,
in some cases (TT~Cyg, X~Her, TZ~Cyg, SW~Gem, U~Hya, S~Sct, SW~Vir)
we could only estimate the periods and amplitudes
with help of a paralel comparison of the multiple structures of the DFT
and the observed cyclic changes in the light curves. The determined
``periods'' in the quoted stars should be rather considered as
characteristic times of variations instead of real periods.

The resulting periodicities can be summarized as follows.
Among the 93 semiregulars, we have found 29 purely
monoperiodic stars, 56 stars with unambiguous multiperiodic
behaviour (44 bi- and 12 triperiodic), and 8 stars which
turned out to be rather irregular (meaning
that we did not find any peak higher than the calculated noise
level for those variables). We present the
main observational properties (average brightness, length of analysed
data in days) as well as the calculated
periods and their amplitudes in Tables 3--5. The period uncertainty
was estimated
from the width of the peaks in the spectra at 90\% of maximum.
One star, RS Cam, apparently has a fourth period ($P_3=81$ days,
$A_3=0.1$ mag.) as well.
The amplitude values have more uncertainty than do the periods
because of the instability of the periods. This issue was studied
by wavelet analysis (see Sect.\ 4 for examples), which is a
useful tool for studying temporal variations in the frequency content
(see, e.g., Bedding et al. 1998, Barth\'es \& Mattei 1997,
Szatm\'ary et al. 1996, Foster 1996, G\'al \& Szatm\'ary 1995a,
Szatm\'ary et al. 1994,
Koll\'ath \& Szeidl 1993, Szatm\'ary \& Vink\'o 1992). Therefore,
the amplitude values listed in Tables 3--5 only serve to indicate
the approximate relative strengths of the corresponding periods.

\begin{table*}
\begin{center}
\caption{Biperiodic variables. The symbols are the same as in Table 3.
Subscripts ``0'' and ``1'' simply correspond to the longer and shorter
periods. Periods being in good agreement with the values listed in the GCVS
are typesetted separately.}
\begin{tabular} {lrrlllll}
\hline
Star    & $\langle m \rangle$ &  $\Delta T$ & P (GCVS)  & $P_0$ & $A_0$ & $P_1$ & $A_1$\\
\hline
ST And  & 10.0   &   5300    & {\bf 328}    & {\bf 338(2)} & 1.15 & 181(1) & 0.20\\
VX And  & 8.5    &   5800    & {\bf 369}    & 904(5) & 0.16 & {\bf 375(2)} & 0.27\\
AQ And  & 8.6    &   8200    & {\bf 346}    & {\bf 346(1)} & 0.15 & 169(1) & 0.18\\
V Aqr   & 8.7    &  22800    & {\bf 244}   & 689(5) & 0.25 & {\bf 241(2)} & 0.35\\
V Aql   & 7.4    &  28300    & 353    & 400(50) & 0.10 & 215(1) & 0.11\\
RS Aur  & 10.0   &  22300    & {\bf 170}    & {\bf 173(1)} & 0.29 & 168(1) & 0.23\\
UU Aur  & 5.7    &  21800    & {\bf 234}   & 441(2)  & 0.15 & {\bf 235(2)} & 0.09\\
V Boo   & 8.7    &  27800    & {\bf 258}    & {\bf 257(1)}  & 0.86 & 137(1) & 0.19\\
RR Cam  & 10.5   &  12400    & {\bf 124}    & 223(2)  & 0.09 & {\bf 124(1)} & 0.10\\
V CVn   & 7.5    &  26400    & {\bf 192}    & {\bf 194(1)}  & 0.42 & 186(1) & 0.13\\
SV Cas  & 8.6    &  25000    & {\bf 265}    & 460(4)  & 0.48 & {\bf 262(2)} & 0.32\\
WZ Cas  & 7.2    &  11000    & {\bf 186}    & 373(1)  & 0.16 & {\bf 187(1)} & 0.09\\
SS Cep  & 7.3    &  27000    & {\bf 90}    & 340(10) & 0.07 & {\bf 100(5)} & 0.05\\
W Cyg   & 6.2    &  33400    & {\bf 131}    & 240(5)  & 0.05 & {\bf 130(5)} & 0.14\\
RS Cyg  & 8.0    &  27800    & {\bf 417}    & {\bf 422(4)}  & 0.48 & 211(2) & 0.20\\
RU Cyg  & 8.5    &  26800    & {\bf 233}    & 441(1)  & 0.16 & {\bf 234(1)} & 0.96\\
RZ Cyg  & 11.8   &  28800    & {\bf 276}    & 537(2)  & 0.63 & {\bf 271(4)} & 0.86\\
TT Cyg  & 8.0    &  25500    & 118    & 390(10) & 0.03 & 188(5) & 0.03\\
TZ Cyg  & 10.8   &   9500    & --    & 138(1)  & 0.06 & 79(1) & 0.05\\
AB Cyg  & 7.9    &  25400    & {\bf 520}    & {\bf 513(2)}  & 0.17 & 429(2) & 0.07\\
AW Cyg  & 8.9    &  22300    & 340    & 3700(50) & 0.10 & 387(3) & 0.10\\
U Del   & 7.0    &  29800    & 110    & 1146(10) & 0.21 & 580(5) & 0.05\\
S Dra   & 8.9    &  26600    & 136    & 311(1) & 0.12 & 172(2) & 0.12\\
RY Dra  & 7.0    &   8800    & 200    & 1150(20) & 0.20 & 300(10) & 0.10\\
UX Dra  & 6.7    &   8000    & {\bf 168}    & 317(2)  & 0.08 & {\bf 176(1)} & 0.10\\
AH Dra  & 7.8    &   8600    & 158    & 189(1) & 0.25 & 107(1) & 0.12\\
RS Gem  & 10.6   &  12400    & {\bf 140}    & 271(1) & 0.25 & {\bf 148(1)} & 0.22\\
X Her   & 6.7    &  33500    & {\bf 95}    & 178(5) & 0.05 & {\bf 102(5)} & 0.03\\
ST Her  & 7.9    &  25000    & {\bf 148}    & 263(2) & 0.08 & {\bf 149(1)} & 0.08\\
g Her   & 5.1    &   9000    & {\bf 89}    & 887(5)  & 0.20 & {\bf 90(1)} & 0.07\\
V Hya   & 9.1    &  32600    & {\bf 531}    & 6400(50) & 1.22 & {\bf 531(3)} & 0.66\\
RY Leo  & 10.2   &  22000    & {\bf 155}    & {\bf 160(1)} & 0.40 & 145(1) & 0.28\\
U LMi   & 11.8   &  10000    & {\bf 272}    & {\bf 272(2)} & 0.45 & 144(1) & 0.16\\
W Ori   & 6.5    &  33300    & {\bf 212}    & 2390(20) & 0.15 & {\bf 208(1)} & 0.08\\
BQ Ori  & 7.9    &  26000    & 110    & 240(6) & 0.14 & 127(2) & 0.10\\
Y Per   & 9.4    &  24000    & {\bf 249}    & {\bf 245(1)} & 0.36 & 127(1) & 0.16\\
S Sct   & 7.3    &  22400    & {\bf 148}    & 269(5) & 0.08 & {\bf 149(2)} & 0.10\\
$\tau^4$ Ser & 6.7  &  5800  & 100       & 1240(10) & 0.10 & 111(1) & 0.09\\
Y Tau   & 7.4     &  27000  & {\bf 242}      & 461(2) & 0.18 & {\bf 242(1)} & 0.18\\
Z UMa   & 7.8     &  31000  & {\bf 196}      & {\bf 195(1)} & 0.33 & 100(1) & 0.05\\
RY UMa  & 7.3     &   9800  & {\bf 310}      & {\bf 305(1)} & 0.16 & 287(1) & 0.11\\
ST UMa  & 6.8     &  26000  & 110      & 5300(20) & 0.11 & 615(6) & 0.09\\
R UMi   & 9.7     &  28000  & {\bf 326}      & {\bf 325(1)} & 0.42 & 170(1) & 0.11\\
RU Vul  & 9.3     &  19500  & 174      & 369(2) & 0.13 & 136(1) & 0.11\\
\hline
\end{tabular}
\end{center}
\end{table*}

\begin{table*}
\begin{center}
\caption{Triply periodic variables.}
\begin{tabular} {lrrlllllll}
\hline
Star    & $\langle m \rangle$ &  $\Delta T$ & P (GCVS) & $P_0$ & $A_0$ & $P_1$ & $A_1$ & $P_2$ & $A_2$\\
\hline
U Cam   & 8.2 & 26800 & -- & 2800(100) & 0.13 & 400(30) & 0.09 & 220(5) & 0.09\\
RS Cam  & 8.7 & 25000 & {\bf 89} & 966(10) & 0.17 & 160(1) & 0.15 & {\bf 90(1)} & 0.12\\
ST Cam  & 7.3 & 28000 & 300 & 1580(10) & 0.10 & 372(3) & 0.12 & 202(2) & 0.08\\
X Cnc   & 6.7 & 25800 & {\bf 195} & 1870(10) & 0.08 & 350(3) & 0.08 & {\bf 193(1)} & 0.09 \\
Y CVn   & 5.7 & 28500 & {\bf 157} & 3000(100) & 0.08 & 273(3) & 0.06 & {\bf 160(2)} & 0.05\\
AF Cyg  & 7.2 & 26600 & {\bf 93} & 921(10) & 0.08 & 163(1) & 0.11 & {\bf 93(1)} & 0.11\\
TX Dra  & 7.6 & 26800 & {\bf 78} & 706(2) & 0.10 & 137(1) & 0.06 & {\bf 77(3)} & 0.07\\
UW Her  & 8.1 & 8600 & {\bf 104} & 1000(10) & 0.09 & 172(1) & 0.08 & {\bf 107(1)} & 0.09\\
Y UMa   & 8.6 & 32000 & {\bf 168} & 324(1) & 0.16 & 315(1) & 0.09 & {\bf 164(2)} & 0.06\\
RX UMa  & 10.6 & 33200 & {\bf 195} & {\bf 201(1)} & 0.37 & 189(1) & 0.26 & 98(0.5) & 0.16\\
V UMi   & 8.1 & 29000 & {\bf 72} & 737(10) & 0.06 & 126(2) & 0.04 & {\bf 73(0.5)} & 0.06\\
SW Vir  & 7.6 & 8500 & {\bf 150} & 1700(50) & 0.15 & 164(1) & 0.13 & {\bf 154(1)} & 0.20\\
\hline
\end{tabular}
\end{center}
\end{table*}

\subsection{Discussion of the multiperiodic nature}

Mattei et al. (1998, hereafter M98) found 30 semiregular variables
with two periods.
Our periods for the 16 common stars
are in very good agreement. This is also true for the two triply
periodic variables, V~UMi and TX~Dra. Since the semiregulars
have quite noisy light curves due to the intrinsic short-timescale
structure, it is worth comparing the independently
determined periodicities by plotting period ratios
against our periods (Fig.\ 7). The significantly deviant points are
those of V~UMi, Y~CVn and S~Dra. This can be explained by the
instability of the periods and the different length of the analysed
data in M98.
The time span of the dataset studied here is more than twice
that of M98. Because the period and amplitude may be changing
over time, the results obtained with datasets covering different
time spans will be different. Thus, we conclude that
the applied period determination and alias filtering
give consistent results with the earlier independent study.

\begin{figure}
\begin{center}
\leavevmode
\psfig{figure=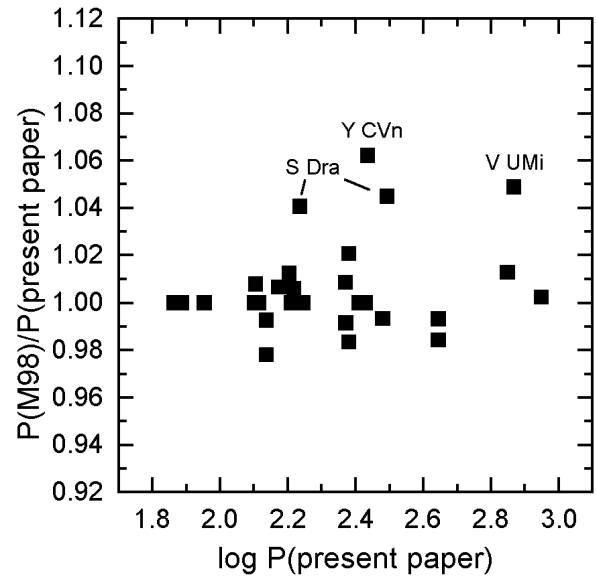,width=\linewidth}
\caption{Intercomparison of 24 periods for 16 stars in common with
Mattei et al. (1998). The differences do not exceed the amount of
instrinsic instability of the periods (up to a few percent cycle-to-cycle
changes).}
\end{center}
\label{f7}
\end{figure}

In order to examine the general distribution of the periods, we
made pairs of periods in 56 multiperiodic variables. This was done
in triply periodic stars
by sorting the periods and choosing the two neighbouring values.
Shorter periods against the longer ones are plotted in Fig.\ 8.
Three sequences are clearly present while two others are suggested.
All of them are marked by dashed lines that were
drawn by fitting a least-squares linear trend to the most
populated sequence and shifting that line to match the other sequences.
It is very interesting how parallel these ridges of data points are.
One would expect
such a separation between stars pulsating in different modes (i.e., for
the same ``longer'' period value different ``shorter'' periods
correspond), assuming that the periodicities are due to pulsation.

\begin{figure}
\begin{center}
\leavevmode
\psfig{figure=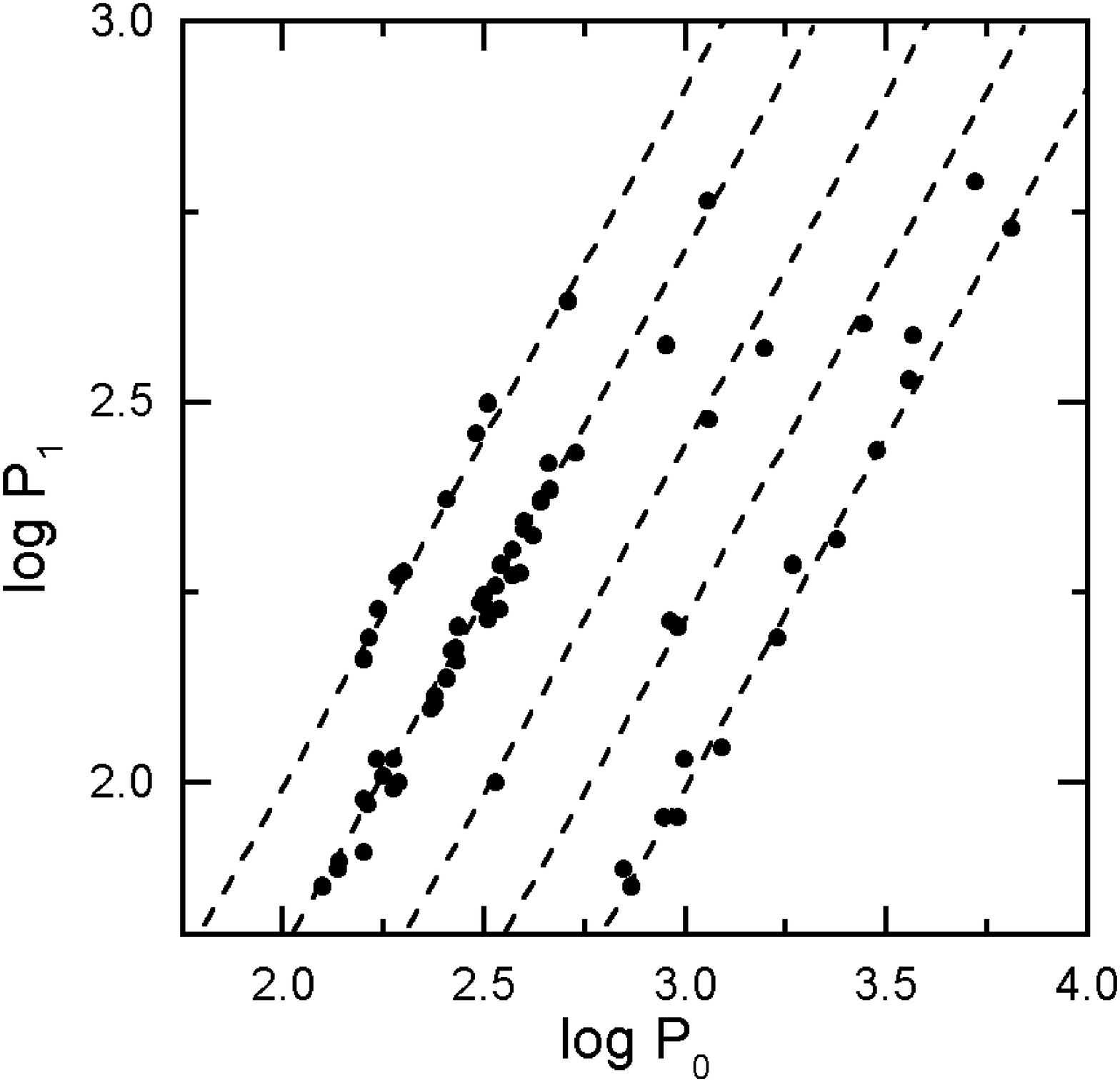,width=\linewidth}
\caption{The ``shorter'' periods vs. the ``longer'' periods of pairs
of periods for the multiply periodic variables. Three sequences are obvious
and two other ones are possible.}
\end{center}
\label{f8}
\end{figure}

Since pulsation theory usually uses period ratios to predict the modes,
we plotted the Petersen diagram (period ratios vs. periods) in Fig.\ 9.
The most populated region contains stars
with period ratios between 1.80 and 2.00, while the other
sequences are those of with ratios of around 10--12, 6.0, 3.60--3.90
and 1.02--1.10. Mattei et al. (1998) pointed
out that in their sample 63\% (19 of 30) of the multiperiodic stars
have period ratios between 1.80 and 2.00.
Our larger sample supports this statistic,
as 39 of 56 stars (70\%) have period ratios of around $1.90\pm0.15$. The
dashed lines in Fig.\ 9 were fitted by the same procedure as in Fig.\ 8.

\begin{figure}
\begin{center}
\leavevmode
\psfig{figure=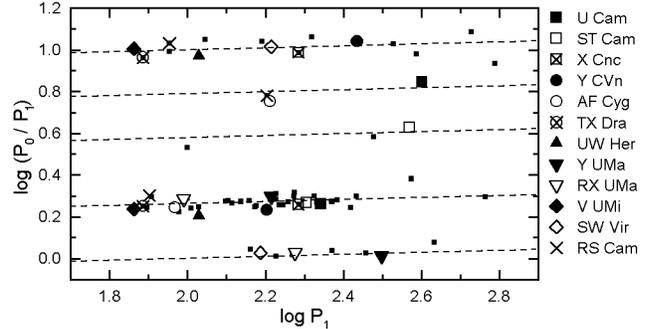,width=\linewidth}
\caption{The Petersen diagram for all multiperiodic semiregulars
with triply-periodic variables shown individually.}
\end{center}
\label{f9}
\end{figure}

The other period ratios are interesting, too. The lowest sequence is that
of stars with period ratios somewhat higher than 1. This means two closely
separated periods, assuming that the close peaks are not due to
small changes of one period. We have checked the original light
curves and found very clear examples for beating (e.g. RU~And, RX~UMa).
On the other hand, some stars should be considered as monoperiodic
variables with slightly and randomly changing period. Unfortunately,
it is very difficult to distinguish between these possibilities.
The upper sequence in Fig.\ 9 is populated by stars with period ratios
of around
10. This is a well-known period ratio for semiregulars
(e.g. Houk 1963, Wood 1976, Percy \& Polano 1998). The intermediate
ratios were not
discussed in the earlier papers, although they are present in their
observational analyses to a smaller extent (see Fig.\ 6 in M98).

The assumption of a few simultaneously excited modes can be tested
by the triply-periodic
variables. If their period ratios fall on the sequences defined
by the doubly periodic stars that would support the assumption.
Fig.\ 9 shows those stars separately. The
main sequences are evidently well covered by the triply-periodic
semiregulars only, too.
Most recently, Bedding et al. (1998) have studied the mode switching
in R~Dor ($P_{long}/P_{short}=1.81$) concluding that
it probably pulsates in the
first and third overtones. Furthermore, they suggest that all stars
with similar period ratios pulsate in these modes. Wood et al. (1998)
presented multiple structures in a diagram similar to Fig.\ 8 for
the LMC red variables, and also suggested higher modes than fundamental
and first overtone.

Although Figs.\ 8--9 supports the idea that the segregation is a consequence
of the presence of many modes of pulsation, other possible explanations
could not be excluded. Older models by Fox \& Wood (1982) predict high
period ratios (6--10) for masses as high as 6--8 $M_{\odot}$, while
the periods of fundamental and first overtone radial modes
have ratios of about 2 in many theoretical models
(e.g. Ostlie \& Cox 1986, Fox \& Wood 1982). On the other hand,
quasi-periodic cycles might be caused by physical mechanisms other
than pulsation (e.g. duplicity, distorted stellar shapes,
rotation -- Barnbaum et al. 1995, dust-shell dynamics --
H\"ofner et al. 1995).
Nevertheless, we can claim that: {\it i)}
a significant percentage
of semiregular stars show multiperiodic behaviour; {\it ii)} there is
supporting evidence provided by the triply periodic variables
that the segregation in Figs.\ 8--9 is due to different modes of pulsation.

We have tried to find correlations among the periods, the period
ratios, and several main physical properties, such as the infrared
JHKL'M colours (Kerschbaum \& Hron 1994), galactic latitude, and
mass-loss rates (Loup et al. 1993). No correlation was
found among these parameters. We have also tried to find
a distinction between the C-rich and O-rich variables, but the
photometric parameters studied did not allow to determine such
a discrimination. Nevertheless, we plotted the period distribution
of the two types of stars in Fig.\ 10. This diagram is strongly
biased by the effects of the sample selection, as noted by the
referee: long-period O-rich stars would have on the average
larger amplitudes (due to the O-rich opacity sources, such as
VO, TiO), and would be classified as Miras and consequently not enter the
sample. Visual C-rich stars with small amplitudes have on
the average higher luminosities and therefore longer periods.
The simple Gaussian fits marked by the solid and
dashed lines were used to estimate the maximum and the spread
of the distributions (186 and 295 days for O-rich and C-rich
stars, respectively; the FWHM is 0.44$\pm$0.07 dex for both fits).

\begin{figure}
\begin{center}
\leavevmode
\psfig{figure=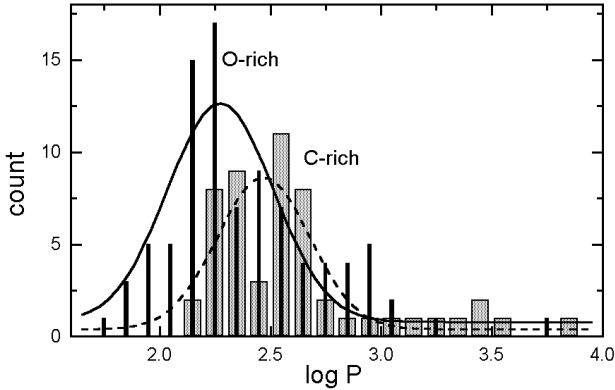,width=\linewidth}
\caption{The period distribution of C-rich and O-rich semiregulars.}
\end{center}
\label{f10}
\end{figure}

\section{Special cases}

This section deals with examples illustrating our analysis procedure and the
application of wavelet analysis. The stars mentioned below as well all
other stars will
be investigated in more details in a subsequent paper. Here we
briefly outline only what we found especially interesting. The examples
cover triple periodicity (TX~Dra and V~UMi), amplitude
modulation (RY~UMa) and long-term amplitude decrease
(V~Boo, RU~Cyg and Y~Per).

\subsection{TX Draconis \& V Ursae Minoris}

\begin{figure}
\begin{center}
\leavevmode
\caption{Fourier spectrum and wavelet contour map for TX Dra. The small
insert shows the window function (the frequency range is
$-$0.015--0.015 c/d). There are a few occasions (indicated
by horizontal arrows) when the 77 days mode (vertical arrow) dominated
the spectrum.}
\end{center}
\label{f11}
\end{figure}

\begin{figure}
\begin{center}
\leavevmode
\caption{The same as in Fig.\ 10 for V UMi. The close similarity is evident.}
\end{center}
\label{f12}
\end{figure}

The clearest examples of triple periodicity are TX~Dra and
V~UMi. In many respects they are twins in their pulsational
characteristics. The dominant modes of TX Dra and V UMi have periods
of 77 days and 73 days, respectively.
The other periods
are also very similar: 706 and 137 days for TX~Dra versus
737 and 126 days for V~UMi. This result is in perfect
agreement with that of Mattei et al. (1998), except for the longest
period in V~UMi. This can be explained by the instability of this
period which caused a double peak around 750 days in the Fourier
spectrum with slightly differing amplitudes. These peaks correspond
exactly to our 737$\pm$10 days and M98's 773 days periods. The data
distribution is not the same in the two analysed data sets which affected
the calculated amplitudes.

We studied the stability of the frequency content by wavelet
analysis (e.g. Szatm\'ary et al. 1996, Foster 1996, Szatm\'ary et al. 1994).
The resulting three-dimensional wavelet contour maps and the
corresponding Fourier spectra are shown in Figs.\ 11-12.
The most unstable period in TX~Dra is the shortest one (77 days). While the
other two modes seem to be quite stable over thousands of days,
the short period component sometimes switches on and off. That is
why there are many peaks in the power spectrum scattering around the
average value of 77 days.
Observers should note that the dominant mode is, as of February 1999,
this rapid one, and based on earlier behaviour, we expect
that it will switch off around 1999--2000, thereby offering
a very good opportunity to observe mode switching in real-time!

V~UMi is generally very similar to TX~Dra, even in the instability
of the excited modes. Obviously the pulsation in these stars
is not a smooth and repetitive process. Mild chaos is probably
present, too, as suggested by, e.g., Mattei et al. (1998). The extended
atmosphere with strong inner convection creates a very complex environment
where slight changes in the actual parameters have very serious
effects on the resulting pulsational properties.
                                                      
The change between pulsational modes has been detected in the case of some
53~Per stars (Smith 1978), in a rapidly oscillating Ap star
(Kreidl et al. 1991), and in F supergiant (Fernie 1983).
Mode switching in red semiregular stars was reported by Cadmus et al. (1991),
G\'al \& Szatm\'ary (1995b), and Percy \& Desjardins (1996).
Further cases of mode switching and models for this phenomenon are discussed
by Bedding et al. (1998).

\subsection{RY Ursae Majoris}

\begin{figure}
\begin{center}
\leavevmode
\psfig{figure=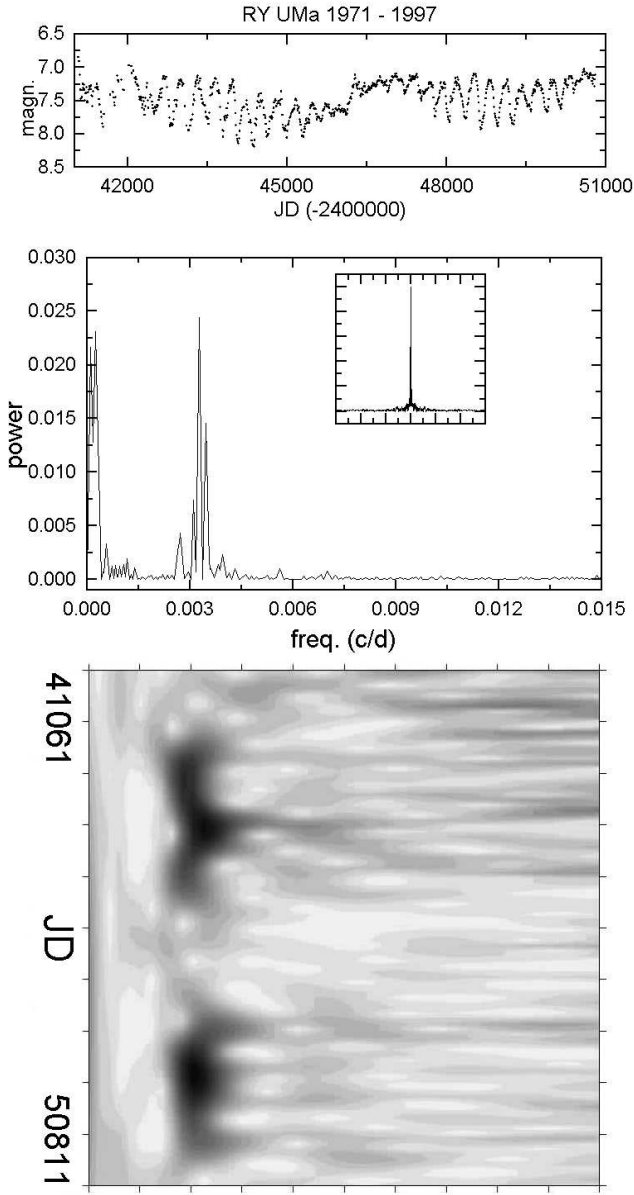,width=\linewidth}
\caption{RY UMa, the best example of amplitude modulation.
The cycle length of the modulation is about 4000 days, which
equals about 13 $P_{puls.}$.}
\end{center}
\label{f13}
\end{figure}

Amplitude modulation in pulsating variables is mainly associated
with RR~Lyrae variables showing the Blazhko-effect (e.g. Kov\'acs 1995,
Szeidl 1988, Moskalik 1986), with $\delta$ Scuti-type stars with
very complex light variations
(e.g. Mantegazza et al. 1996, Breger 1993), one
known classical Cepheid, V473~Lyr, which has strong amplitude
modulation (Van Hoolst \& Waelkens 1995), and some
Mira and semiregular variables (Mattei 1993, Mattei et al. 1998,
Mattei \& Foster 1999a, b, Barth\'es \& Mattei 1997).

\begin{figure*}
\begin{center}
\leavevmode
\caption{Compressed light curves, power spectra and wavelet maps
for V Boo and
RU Cyg. These stars probably evolve from the Mira state to the semiregular
state.}
\end{center}
\label{f14}
\end{figure*}

In our sample, one of the best examples of semiregular variables
with amplitude modulation is RY~UMa, which is classified as
an SRb star. Its light curve reveals a clear amplitude
modulation which is strongly supported by the results of wavelet
analysis (Fig.\ 13). Although the Fourier spectrum suggests two closely
separated periods (see Table 4), accepting them would be misleading, as
the wavelet map shows slight frequency changes of the principal peak
accompanied with amplitude modulation. Therefore, that close
peak is an artifact caused by the instability of the principal
one. The underlying physical mechanism in unknown: there are several
possibilities such as the rotation, magnetic activity change or
duplicity effects. Unfortunately the presently available observations
do not allow finding a reliable model for this modulation. Nevertheless,
it is very interesting that the characteristic time of the
amplitude modulation is about 4000 days being around a typical
value of theoretically calculated rate of rotation of red giant
stars (as estimated using rotational velocities studied by
Schrijver \& Pols 1993).

\subsection{V~Bootis, RU~Cygni \& Y~Persei}

V~Boo is the prototype of SRa variables suffering from long-term amplitude
decrease (Szatm\'ary et al. 1996). Recently Bedding et al. (1998)
presented a very similar phenomenon for R~Dor which was classified in GCVS
as an SRb star. The proposed explanation for the amplitude decrease in R~Dor
is that the star is evolving from the Mira state to the semiregular state.
We found another two examples of amplitude decrease in
RU~Cyg (SRa) and Y~Per (Mira), which are consequently the third and
fourth candidates for that interesting evolutionary status.
Unfortunately, these substantial changes of the lightcurves are only
hints of probable change of the variability type, further spectroscopic
or near-infrared photometric observations would be desirable.
V~Boo and RU~Cyg are compared in Fig.\ 14, where long-term
light curves are plotted together with
the corresponding power spectra and wavelet maps. The similarity is quite
conspicuous.

Y~Per differs from V~Boo and RU~Cyg in a very important aspect.
While the dominant frequencies of V~Boo and RU~Cyg did not change
significantly, Y~Per seems to be a clear example of a transition
from Mira to SRb. This is presented in Fig.\ 15, where the
compressed light curve is plotted with two power spectra
corresponding to two data subsets (before and after JD 47000).
The earlier monoperiodicity (P=253 day) was replaced by
a biperiodicity (P$_{0}$=245 day, P$_{1}$=127 day). Apparently
the Mira star Y~Per was transformed to a typical doubly-periodic SRb star.
The most surprising result is the abruptness
of the mode switching.

\section{Summary}

\begin{figure}
\begin{center}
\leavevmode
\psfig{figure=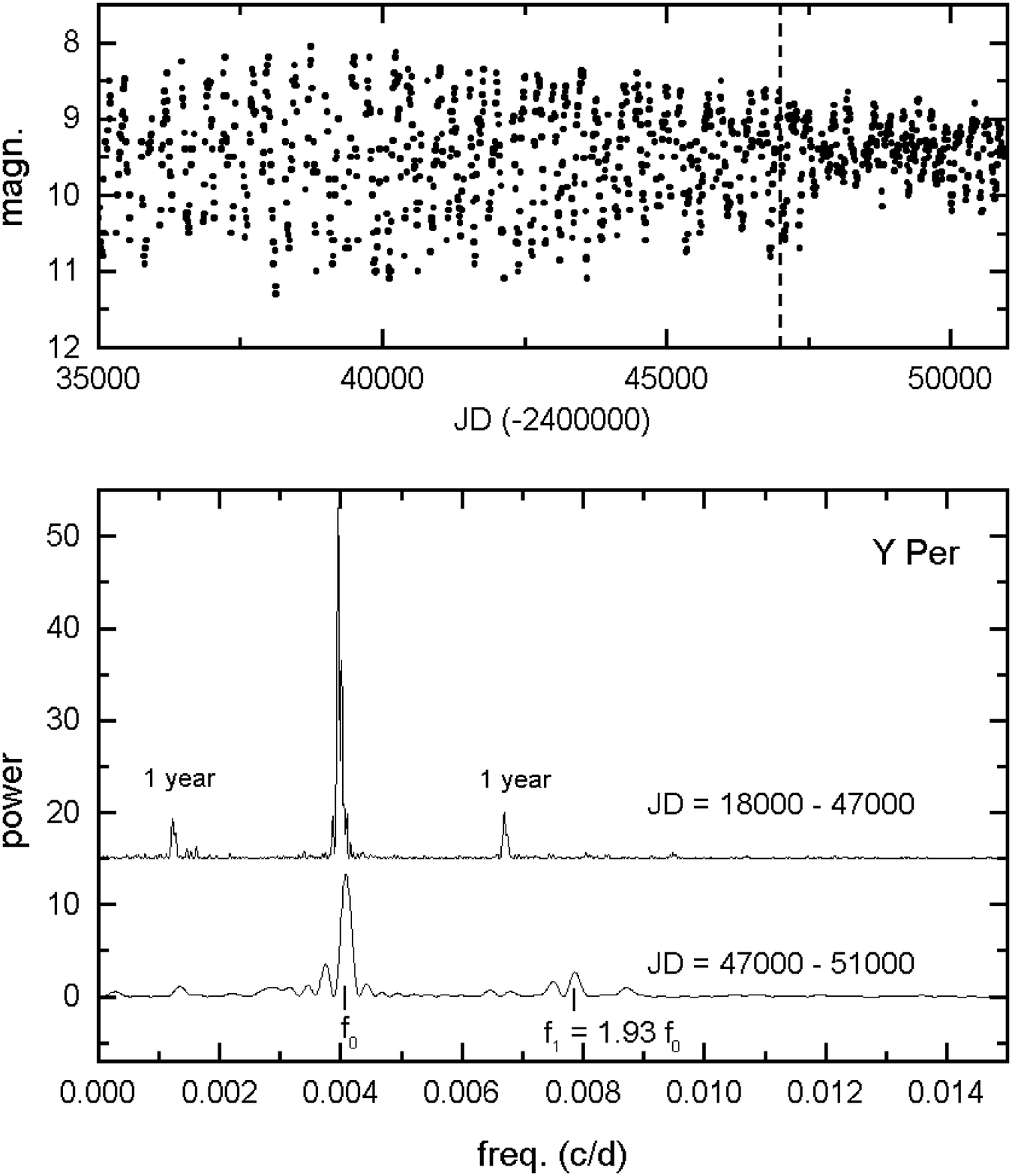,width=\linewidth}
\caption{Y~Per: compressed light curve and power spectra of two subsets
separated by the dashed line.}
\end{center}
\label{f15}
\end{figure}

Based on the light curve analyses presented in the previous sections
we have obtained the following results:

1. We have analysed long-term visual observations of 93 red semiregular
variables in order to determine their dominant periods. Direct comparison
with photoelectric measurements demonstrated the
usefulness of the low quality visual data. The most important requirement
of precise frequency determination is to have as long a time-series
as possible, as has been illustrated by frequency analysis of
artificial data.

2. We have found 29 monoperiodic and 56 multiperiodic stars
(44 with two and 12 with three significant and essentially stable periods).
8 variables do not show any unambiguous periodicity. The distribution
of periods and period ratios in multiperiodic variables suggests
the existence of up to five different groups among the variables studied,
which is most probably due to different modes of pulsation.

3. We have highlighted a few interesting special cases:

\begin{itemize}
\item{} TX~Dra and V~UMi are very similar triply-periodic variables
with nearly equal periods. While the longer two periods are
stable over decades of time, the high-frequency mode switches on and off
from time to time. We predict another mode change of
TX~Dra soon, most probably in 1999-2000.

\item{} RY~UMa is one of the best documented examples for amplitude
modulation in SRV's. Unfortunately the visual data alone are not enough
to determine the underlying physical process.

\item{} We have discussed three stars (V~Boo, RU~Cyg and Y~Per)
that show a gradual decrease in amplitude. All of them seem to evolve from
mira-like to semiregular type (as does R~Dor, according to
Bedding et al. 1998), which suggests that Miras and
SRV's may be much more closely related than was thought earlier.
Y~Per differs from the other stars, because, in addition to the amplitude
decrease, a new mode has appeared with a quite high amplitude.
The observed period ratio (1.93) is very typical in the majority
of doubly periodic SRV's.

\end{itemize}

\begin{acknowledgements}
We sincerely thank variable star observers of AFOEV, VSOLJ, HAA/VSS and
AAVSO whose dedicated observations over many decades made this
study possible. The referee (Dr. F. Kerschbaum) has greatly
improved the paper with his notes and suggestions.
This research was supported by Hungarian OTKA Grants \#F022249,
\#T022259 and Szeged Observatory Foundation. The NASA ADS Abstract
Service was used to access data and references.
\end{acknowledgements}

\end{document}